\documentclass[preprint,aps]{revtex4}
\usepackage{epsf}
\usepackage{graphicx} 
\usepackage{color} 

\begin{document}

\title
{
High-Order Coupled Cluster Method (CCM) Formalism 1: \\
Ground- and Excited-State Properties of \\
Lattice Quantum Spin Systems with $s \ge \frac 12$
}

\author
{
D. J. J. Farnell
}
\affiliation
{Health Methodology Research Group, School of Community-Based Medicine, 
Jean McFarlane Building, University Place, University of Manchester,
Manchester M13 9PL, United Kingdom}

\date{\today}

\begin{abstract}
The coupled cluster method (CCM) is a powerful and widely applied technique of modern-day quantum many-body theory. It has been used with great success in order to understand the properties of quantum magnets at zero temperature. This is due largely to the application of computational techniques that allow the method to be applied to high orders of approximation using localised approximation schemes, e.g., such as the LSUB$m$ scheme. In this article, the high-order CCM formalism for the ground and excited states of quantum magnetic systems are extended to those with spin quantum number $s \ge  \frac 12$. Solution strategies for the ket- and bra-state equations are also considered. Aspects of extrapolation of CCM expectation values are discussed and future topics regarding extrapolations are presented.
\end{abstract}

\maketitle

\section{Introduction}

The coupled cluster method (CCM) \cite{refc1,refc2,refc3,refc4,refc5,refc6,refc7,refc8,refc9}  is a well-known method of quantum many-body theory (QMBT). The CCM has been applied with much success over the last fifteen or so years in order to study quantum magnetic systems at zero temperature (see Refs. \cite{ccm1,ccm2,ccm999,ccm3,ccm4,ccm5,ccm6,ccm7,ccm8,ccm9,ccm10,ccm11,ccm12,ccm13,ccm13a,ccm14,ccm15,ccm16,ccm17,ccm18,ccm19,ccm19a,ccm20,ccm21,ccm22,ccm23,ccm,ccm24,ccm24a,ccm26,ccm27,ccm27a,ccm28,ccm29,ccm30,ccm31,ccm32,ccm33,ccm34,ccm35,ccm36,ccm37,ccm38}  ). In particular, the use of computer-algebraic implementations \cite{ccm12,ccm15,ccm20} of the CCM for quantum systems of large or infinite numbers of particles has largely been found to be very effective with respect to these spin-lattice problems. This approach uses localised approximation schemes, such as the LSUB$m$ approximation. For the LSUB$m$ scheme, the extent of the locale over which multi-spin correlations are explicitly included in the approximation is defined by the index $m$. The ground- and excited-state expectation values are often extrapolated in the limit $m \rightarrow \infty$. In this article we focus on the development of new high-order CCM formalism for the ground and excited states of lattice quantum spin systems with spin quantum number $s \ge \frac 12$. The solution the ket- and bra-state equations is also considered. Various aspects of the extrapolation of CCM expectation values are considered and future topics regarding extrapolations are described. The high-order CCCM code is freely available online \cite{ccm}.

\section{CCM Ground-State Formalism}

The ket and bra ground-state energy eigenvectors, 
$|\Psi\rangle$ and  $\langle\tilde{\Psi}|$, of a general many-body system described 
by a Hamiltonian $H$, are given by
%%%%%%%%%%%%%%%%%%%%%%%%%%%%%%%%%%
\begin{equation} 
H |\Psi\rangle = E_g |\Psi\rangle
\;; 
\;\;\;  
\langle\tilde{\Psi}| H = E_g \langle\tilde{\Psi}| 
\;. 
\label{eq1} 
\end{equation} 
%%%%%%%%%%%%%%%%%%%%%%%%%%%%%%%%%%  
Furthermore, the ket and bra states are parametrised within the single-reference CCM as follows:   
%%%%%%%%%%%%%%%%%% 
\begin{eqnarray} 
|\Psi\rangle = {\rm e}^S |\Phi\rangle \; &;&  
\;\;\; S=\sum_{I \neq 0} {\cal S}_I C_I^{+}  \nonumber \; , \\ 
\langle\tilde{\Psi}| = \langle\Phi| \tilde{S} {\rm e}^{-S} \; &;& 
\;\;\; \tilde{S} =1 + \sum_{I \neq 0} \tilde{{\cal S}}_I C_I^{-} \; .  
\label{eq2} 
\end{eqnarray} 
%%%%%%%%%%%%%%%%%% 
It may be proven from
Eqs. (\ref{eq1}) and (\ref{eq2}) in a straightforward manner that the ket- and bra-state 
equations are thus given by
%%%%%%%%%%%%%%%%% 
\begin{eqnarray} 
\langle\Phi|C_I^{-} {\rm e}^{-S} H {\rm e}^S|\Phi\rangle &=& 0 ,  \;\; 
\forall I \neq 0 \;\; ; \label{eq7} \\ 
\langle\Phi|\tilde{S} {\rm e}^{-S} [H,C_I^{+}] {\rm e}^S|\Phi\rangle 
&=& 0 , \;\; \forall I \neq 0 \;\; . \label{eq8}
\end{eqnarray}  
%%%%%%%%%%%%%%%% 
The index $I$ refers to a particular choice of cluster from the set of 
($N_F$) fundamental clusters that are distinct under the symmetries 
of the crystallographic lattice and the Hamiltonian and for a given 
approximation scheme at a given level of approximation.
We note that these equations are equivalent to the minimization 
of the expectation value of $\bar H = \langle \tilde \Psi | H | \Psi \rangle$ 
with respect to the the CCM bra- and ket-state correlation coefficients 
$\{ \tilde{{\cal S}}_I, {\cal S}_I \}$. We note that Eq. (\ref{eq7}) is equivalent
to $\delta{\bar{H}} / \delta{\tilde{{\cal S}}_I}=0$, whereas Eq. (\ref{eq8}) is 
equivalent to $\delta{\bar{H}} / \delta{{\cal S}_I}=0$. Furthermore, we note 
that Eq. (\ref{eq7}) leads directly to simple form for the ground-state
energy given by
%%%%%%%%%%%%%%%%
\begin{equation} 
E_g = E_g ( \{{\cal S}_I\} ) = \langle\Phi| {\rm e}^{-S} H {\rm e}^S|\Phi\rangle
\;\; . 
\label{eq9}
\end{equation}  
%%%%%%%%%%%%%%%% 
The full set $\{ {\cal S}_I, \tilde{{\cal S}}_I \}$ provides a complete 
description of the ground state. For instance, an arbitrary 
operator $A$ will have a ground-state expectation value given as 
%%%%%%%%%%%%%%%%
\begin{equation} 
\bar{A}
\equiv \langle\tilde{\Psi}\vert A \vert\Psi\rangle
=\langle\Phi | \tilde{S} {\rm e}^{-S} A {\rm e}^S | \Phi\rangle
=\bar{A}\left( \{ {\cal S}_I,\tilde{{\cal S}}_I \} \right) 
\; .
\label{eq6}
\end{equation} 
%%%%%%%%%%%%%%%% 
The similarity transform of $A$ is given by,
%%%%%%%%%%%%%%%%%%%%%%
\begin{equation}  
\tilde A \equiv {\rm e}^{-S} A {\rm e}^{S} = A 
+ [A,S] + {1\over2!} [[A,S],S] + \cdots 
\;\; .
\label{eq10}
\end{equation} 
%%%%%%%%%%%%%%%%%%%%%%
Finally, we remark that the CCM provides exact results in the limit of inclusion of all possible clusters in $S$ and $\tilde S$. However, this problem is often impossible to solve in a practical sense. 
Hence, we generally make approximations in both $S$ and $\tilde S$.  The three most commonly employed approximation schemes previously utilised have been: (1) the SUB$n$ scheme, in which all correlations involving only $n$ or fewer spins are retained, but no further restriction is made concerning their spatial separation on the lattice; (2) the SUB$n$-$m$  sub-approximation, in which all SUB$n$ correlations spanning a range of no more than $m$ adjacent lattice sites are retained; and (3) the localised LSUB$m$ scheme, in which all multi-spin correlations over all distinct locales on the lattice defined by $m$ or fewer contiguous sites are retained. 

\section{High-Order Ground-State Operators and Commutations}

We begin the treatment of high-order CCM by introducing the ket-state correlation operator given, as usual, by 
\begin{equation}
S=\sum_{l} ~ \sum_{i_1,\cdots,i_l} {\cal S}_{ i_1, \cdot\cdot\cdot, i_{l}} 
s^+_{i_1} \cdot\cdot\cdot s^+_{i_{l}} ~~ .
\end{equation}
However, it is important point to note that each of the indices $\{i_1, i_2,\cdots,i_l \}$  runs over {\it all} lattice sites. Furthermore, we assume that there are $(l!)$ orderings of these indices (even for $s > \frac 12$), although we never need to work out these factors explicitly in practice. 
The index $I$ corresponds to one of the choices of $\{i_1, \cdots, i_l \}$ for the
fundamental set of configurations. 
We may now write a set of high-order CCM ket-state operators, given by
\begin{equation}
~ \left .
\mbox{
\begin{tabular}{l@{~}l@{~}l@{~~}}
$F_{k}$   &$\equiv$ &$\sum_{l} ~ \sum_{i_2,\cdots,i_l} ~ 
l {\cal S}_{k, i_2, \cdot\cdot\cdot, i_{l}} ~
s^+_{i_2} \cdot\cdot\cdot s^+_{i_{l}}$\\
$G_{km}$  &$\equiv$ &$\sum_{l>1} ~ \sum_{i_3,\cdots,i_l} ~ 
l(l-1) {\cal S}_{k, m, i_3, \cdot\cdot\cdot, i_{l}} ~
s^+_{i_3} \cdot\cdot\cdot s^+_{i_{l}}$  \\
$M_{kmn}$ &$\equiv$ &$\sum_{l>2} ~ \sum_{i_4,\cdots,i_l} ~ 
l(l-1)(l-2){\cal S}_{k, m, n, i_4, \cdot\cdot\cdot, i_{l}} ~
s^+_{i_4} \cdot\cdot\cdot s^+_{i_{l}}$  \\
$N_{kmnp}$&$\equiv$ &$\sum_{l>3} ~ \sum_{i_5,\cdots,i_l} ~
l(l-1)(l-2)(l-3){\cal S}_{k, m, n, p, i_5, \cdot\cdot\cdot, i_{l}} ~
s^+_{i_5} \cdot\cdot\cdot s^+_{i_{l}}$ 
\end{tabular}
}
\right \}
\label{eq13}
\end{equation}
The indices $k$, $m$, $n$, and $p$ depend on those sums in the Hamiltonian or 
of another given operator. We note that $s^\pm = s^x \pm {\rm i}s^y$, 
$[s^z,s^{\pm}]=\pm s^{\pm}$,  and $[s^-,s^+]=-2 s^z$. 
Hence, the following commutation relations may be proven:
\begin{equation}
~ \left .
\mbox{
\begin{tabular}{l@{~}l@{~}l@{~~}}
$[s_k^{z},S]$    ~ &=& ~ $F_{k} s_k^+$~~,\\
$[s_k^{-},S]$  ~ &=& ~ $-2F_{k} s_k^z - G_{kk}s_k^+$~~,\\
$[s_k^{z},F_{m}]$    ~ &=& ~ $G_{km} s_k^+$~~,\\
$[s_k^{z},G_{mn}]$  ~ &=& ~ $M_{kmn} s_k^+$~~,\\
$[s_k^{z},F_m^2]$    ~ &=& ~ $2 F_m G_{km} s_k^+$~~,\\
$[s_k^{-},F_{m}]$    ~ &=& ~ $-2 G_{km} s_k^z - M_{kkm} s_k^+$~~,\\
$[s_k^{-},F_{m}^2]$  ~ &=& ~ $-2 G_{km}^2 s_k^+ - 2 F_m M_{kkm} s_k^+
-4F_m G_{km} s_k^z $~~,\\
$[s_k^{z},M_{mnp}]$  ~ &=& ~ $N_{kmnp} s_k^+$~~,\\
$[s_k^{-},G_{mn}]$  ~ &=& ~ $-2 M_{kmn} s_k^z - N_{kkmn} s_k^+$~~.\\
\end{tabular}
}
\right \} ~~
\label{appendix9}
\end{equation}
We may now write the similarity-transformed expressions
of the single-spin operators $s^\alpha ~ ; ~ \alpha\equiv\{+,-,z\}$,
as 
\begin{equation}
~ \left .
\mbox{
\begin{tabular}{l@{~}l@{~}l@{~}l@{~~}}
$e^{-S} s_k^+ e^{S} \equiv$ &$\tilde s_k^+$  &$=$&  $s_k^+$   \\
$e^{-S} s_k^z e^{S} \equiv$ &$\tilde s_k^z$  &$=$&  $s_k^z + F_k s_k^+$   \\
$e^{-S} s_k^- e^{S} \equiv$ &$\tilde s_k^-$  &$=$&  
$s_k^- -2 F_k s_k^z - G_{kk} s_k^+- F_k^2 s_k^+$ ~~ .\\
\end{tabular}
}
\right \}
\label{eq14}
\end{equation} 
We see that there is  a repeated index in $G_{kk}$ in the similarity transformed version 
of $s^-$. Clearly, this term contributes only for systems with spin quantum number 
$s > \frac 12$. 

\section{Deriving and Solving The CCM Ground-State Equations}

We now wish to determine and solve the CCM ket-state equations, where the 
$I$-th such equation is given by
\begin{equation}
E_I \equiv  \langle \Phi | C_I^- e^{-S} H e^{S} | \Phi \rangle = 0 ~~,
\forall I \ne 0 ~~ .
\label{tempLabel}
\end{equation}
(Note that we assume that $\langle \Phi | C_I^- C_I^+ | \Phi \rangle = 1$ in the above
equation). 
Specific terms in the Hamiltonian are now explicitly written in terms of the high-order CCM
operators as:
\begin{eqnarray}
{\rm TERM~1:} ~ \tilde s^z_i \tilde s^z_j &=& 
s^z_i  s^z_j + F_j s_j^+ s_i^z + F_i s_i^+ s_j^z + 
G_{ij} s_i^+ s_j^+ + F_i F_j  s_i^+ s_j^+ \nonumber \\
{\rm TERM~2:} ~ \tilde s^z_i \tilde s^+_j &=& 
s^+_j  s^z_i + F_i s_i^+ s_j^+ 
\nonumber \\
{\rm TERM~3:} ~ \tilde s^z_i \tilde s^-_j &=& 
-2 F_j s_i^z s_j^z - 2 G_{ij} s_i^+ s_j^z - G_{jj}  s_j^+ s_i^z -M_{ijj}  s_i^+ s_j^+ 
-2 F_i F_j s^+_i  s^z_j \nonumber \\
&&-2F_j G_{ij} s_i^+ s_j^+  - F_i G_{jj} s_i^+ s_j^+ - F_i F_j^2 s_i^+ s_j^+ 
-F_j^2 s_j^+ s_i^z
\nonumber \\
{\rm TERM~4:} ~ \tilde s^+_i \tilde s^z_j &=& 
s^+_i  s^z_j + F_j s_i^+ s_j^+ 
\nonumber \\
{\rm TERM~5:} ~ \tilde s^-_i \tilde s^z_j &=& 
-2 F_i s_i^z s_j^z - 2 G_{ij} s_j^+ s_i^z - G_{ii}  s_i^+ s_j^z -M_{iij}  s_i^+ s_j^+ 
-2 F_i F_j s^+_j  s^z_i \nonumber \\
&&-2F_i G_{ij} s_i^+ s_j^+  - F_j G_{ii} s_i^+ s_j^+ - F_j F_i^2 s_i^+ s_j^+ 
-F_i^2 s_i^+ s_j^z\nonumber \\
{\rm TERM~6:} ~ \tilde s^+_i \tilde s^-_j &=& 
-2 F_j s_i^+ s_j^z - G_{jj} s_i^+ s_j^+ -F_j^2 s_i^+ s_j^+
\nonumber \\
{\rm TERM~7:} ~ \tilde s^-_i \tilde s^+_j &=& 
-2 F_i s_j^+ s_i^z - G_{ii} s_i^+ s_j^+ -F_i^2 s_i^+ s_j^+
\nonumber \\
{\rm TERM~8:} ~ \tilde s^+_i \tilde s^+_j &=& 
s_i^+ s_j^+
\nonumber \\
{\rm TERM~9:} ~ \tilde s^-_i \tilde s^-_j &=& 
4G_{ij} s_i^z s_j^z + 2 M_{iij} s_i^+ s_j^z + 2 M_{ijj} s_j^+ s_i^z + N_{iijj} s_i^+ s_j^+
\nonumber \\
&&+2G_{ij}^2 s_i^+ s_j^+ + 2 F_j M_{iij} s_i^+ s_j^+ + 4 F_j G_{ij} s_j^+ s_i^z + 4F_i F_j s_i^z s_j^z
\nonumber \\
&&+4F_iG_{ij}  s_i^+ s_j^z + 2 F_i G_{jj} s_j^+ s_i^z + 2  F_i M_{ijj} s_i^+ s_j^+ + 4 F_i F_j G_{ij} 
s_i^+ s_j^+ 
\nonumber \\
&&+2F_i F_j^2 s_j^+ s_i^z + 2 F_j G_{ii} s_i^+ s_j^z + G_{ii} G_{jj} s_i^+ s_j^+ + F_j^2 G_{ii}  s_i^+ s_j^+
\nonumber \\
&&+2F_i^2 F_j s_i^+ s_j^z +F_i^2 G_{jj}  s_i^+ s_j^+ + F_i^2 F_j^2  s_i^+ s_j^+
\nonumber \\
{\rm TERM~10:} ~ \tilde s^z_i &=& 
s_i^z + F_i s_i^+
\nonumber \\
{\rm TERM~11:} ~ \tilde s^-_i &=& 
-2 F_i s_i^z - G_{ii} s_i^+- (F_i)^2 s_i^+
\nonumber \\
{\rm TERM~12:} ~ (\tilde {s}^z_i)^2 &=& 
(s_i^z)^2 + 2F_i s_i^+ s_i^z + G_{ii} (s_i^+)^2 + F_i (s_i^+)^2 + F_i^2 (s_i^+)^2 
\nonumber \\
{\rm TERM~13:} ~ \tilde s^+_i &=& 
s_i^+
\label{gs}
\end{eqnarray}
(Note that $s^-|\Phi\rangle=0$ is implicitly assumed in Eq. (\ref{gs}) above.) 
We now ``pattern-match''  the $C_i^-$ operators to those the relevant terms in the 
Hamiltonian from Eq. (\ref{gs}) above in order to form the CCM equations $E_I =0$ 
of Eq. (\ref{tempLabel}) at a given level of approximation. 
%However, these are solved via direct iteration for larger values of the approximation
%level because the cost of storing the Jacobian in local memory for the Newton-Raphson 
%(or other) method becomes prohibitive. This may be parallelized also to achieve
%very high orders of approximation and this is discussed below. 

We now define the following {\it new} set of CCM bra-state correlation 
coefficients given by  $x_I\equiv {\cal S}_I$  and $\tilde{x}_I\equiv 
N_B/N (l!)\nu_I \tilde{\cal S}_I$  and we assume again that 
$ \langle \Phi | C_I^- C_I^+ | \Phi \rangle=1$. 
Note that $N_B$ is the number of Bravais lattice sites. Note also 
that for a given cluster $I$ then $\nu_I$ is a symmetry factor which 
is dependent purely on the point-group symmetries (and {\it not} 
the translational symmetries) of the crystallographic lattice and 
that $l$ is the number of spin operators. We note 
that the factors $\nu_I$, $N$, $N_B$, and $(l!)$ never need 
to be explicitly determined. 
The CCM bra-state operator may thus be rewritten as
\begin{equation}
\tilde S \equiv 1 + N \sum_{I=1}^{N_F} {\tilde x_I} C_I^- ~~ ,
\end{equation}
such that we have a particularly simple form for $\bar H$, given by
\begin{equation}
\bar{H} = N \sum_{I=0}^{N_F} \tilde x_I E_I ~~,
\label{appendix15}
\end{equation}
where $\tilde x_0=1$. 
We note that the $E_0$ is
defined by $E_0 = \frac 1N \langle \Phi| e^{-S} H e^S | \Phi \rangle$ 
(and, thus, $E_0=\frac 1N E_g$) 
and that $E_I$ is the $I$-th CCM ket-state equation defined
by Eq. (\ref{tempLabel}). The CCM ket-state equations are easily 
re-derived by taking the partial derivative of $\bar{H}/N$ with 
respect to $\tilde x_I$, where
\begin{equation}
\frac {\delta{(\bar{H}/N)}}{\delta \tilde x_I} (\equiv 0) = E_I ~~.
\label{appendix16}
\end{equation}
We now take the partial derivative of $\bar{H}/N$ with respect to 
$x_I$ such that the bra-state equations take on a particularly 
simple form, given by
\begin{equation}
\frac {\delta{(\bar{H}/N)}}{\delta x_I} =
\frac {\delta{E_0}}{\delta x_I} + 
\sum_{J=1}^{N_F} \tilde x_J \frac {\delta{E_J}}{\delta x_I} (\equiv 0) = \tilde E_I~~.
\label{appendix17}
\end{equation}
The coupled non-linear equations for the ket state $E_I =0$ are solved readily, e.g,  
by using the Newton-Raphson method, in order to find the coefficients $\{x_I \}$.  
By contrast, the equations for the bra state $\tilde E_I =0$ are easily 
solved via LU decomposition,
% for low to medium orders of approximation or via direct iteration (which may be parallelized,  
%as discussed below) for even higher orders of  approximation.
although this may only be carried out once the CCM ket-state
equations have been determined and solved. The numerical values 
of the coefficients $\{\tilde x_I \}$ may thus be obtained. We note
that this approach greatly simplifies the task of determining the
bra-state equations because we infer the bra-state equations directly 
from those of the ket-state equations via Eq. (\ref{appendix17}). Thus, 
we never need to evaluate Eq. (\ref{eq8}) explicitly. 

We may also solve the ket- and bra-state equations (i.e., $E_I=0$ 
and $\tilde E_I=0$, respectively) via direct iteration. For the case of the ket-state 
equations this is slightly more complicated because there are non-linear 
terms with respect the ket-state correlation coefficients $\{ x_I \}$. We rearrange
the ket-state equations such that the linear terms for $x_I$ for the $i^{{\rm th}}$ 
CCM ket-state equation are on the left of the new equation and all other terms are 
on the right. The right-hand side of this new equation is denoted by $E_I'$ after
dividing through by the factor on the left-hand-side for the ket-state correlation
coefficients. We may carry out exactly the same procedure for the bra-state
in order to find $\tilde E_I'$, although the problem is linear with respect to 
$\{ \tilde x_I \}$  in this case. These equations are thus rewritten 
conveniently for the ket state as
\begin{eqnarray}
x_I &=& E_I'(x_1, x_2, \cdots , x_{I-1}, x_{I+1}, \cdots , x_{N_F}, x_1^2, x_2^2, \cdots, x_1^4, \cdots, x_{N_F}^4) ~~ ,
\label{direct1}
\end{eqnarray}
and for the bra state as,
\begin{eqnarray}
\tilde x_I &=& \tilde E_I'(\tilde x_1, \tilde x_2, \cdots , \tilde x_{I-1}, \tilde x_{I+1}, \cdots , \tilde x_{N_F} ~ ; ~
x_1, x_2, \cdots , x_{N_F}, x_1^2, x_2^2, \cdots, x_1^3, \cdots, x_{N_F}^3) ~~.
\label{direct2}
\end{eqnarray}
Clearly, these equations may be solved for $x_I$ and $\tilde x_I$ by iterating them
``directly'' until convergence. Indeed, the local memory usage is vastly reduced because we
do not need to store any Jacobian or other large matrix that scales in size with $N_F^2$.  
This simple ``brute force'' approach of direct iteration has actually been found to be surprisingly 
successful. However, in practice, it needs to be implemented for large numbers of
CPUs used in parallel for very large numbers of clusters (e.g., $10^6$) used in $S$ and 
$\tilde S$. Clearly, more sophisticated solvers for the bra- and ket-state equations that do 
not demand the memory requirements of Newton-Raphson for the ket state and LU 
decomposition for the bra state and are quicker than direct iteration may be implemented. 
However, this remains a task for the future.

%Furthermore, the computational problem posed by solving Eqs. (\ref{direct1}) and (\ref{direct2}) 
%via direct iteration may be solved using parallel processing. The different equations
%of Eqs. (\ref{direct1}) and (\ref{direct2}) for different values of the index $I$ are
%determined separately on different processors. The resulting data for these equations 
%for the different values of $I$ are then stored locally to each processor. 
%As each iteration of the ``direct iteration'' 
%method we find the right-hand sides of those relevant values of $I$ allocated to each 
%processor. We then collect the right-hand side into a single array and this forms our 
%values for $x_I$ or $\tilde x_I$ for the next iteration. Again, we note that we must solve
%the ket-state equations of Eq. (\ref{direct1}) first and then these values for the ket-state 
%coefficients are used in the bra-state equations of Eq. (\ref{direct2}).  This approach is a simple 
%``brute-force''  method, although it has been found to be surprisingly successful at going to very
%high orders of approximation. Indeed, we may now treat of order $10^6$ fundamental 
%clusters using this approach and for  approximately $10^2-10^3$ processors used in parallel.
%Clearly, a similar approach may also be used to find the ``generalized'' expectation values
%via parallel processing. 

\section{The Excited-State Formalism}

We now consider how the excited state may be treated using the CCM via
a high-order approach. We begin by remarking that the excited-state wave function 
is given by
\begin{equation}
|\Psi_e\rangle = X^e ~ e^S |\Phi\rangle ~~ .
\label{eq16}
\end{equation} 
The Schr{\"o}dinger equation, $E_e |\Psi_e\rangle = 
H  |\Psi_e\rangle$ and the equivalent equation for the ground
state lead (after some simple algebra) to 
\begin{equation}
\epsilon_e X^e  | \Phi \rangle = e^{-S} [H,X^e] e^S | 
\Phi \rangle ~ (\equiv  \hat R | \Phi \rangle) ~~ ,
\label{eq17}
\end{equation}
where $\epsilon_e \equiv E_e-E_g$ is the excitation energy. 
We note that the excited-state correlation operator is written as,
\begin{equation}
X^e = \sum_{I \ne 0} {\cal X}_I^e C_{I}^+ ~~ ,
\label{eq18}
\end{equation}
Equation (\ref{eq18}) implies the overlap relation 
\begin{eqnarray}
\langle \Phi | \Psi_e \rangle &=& \langle \Phi | X^e e^S | \Phi \rangle \nonumber \\
\Rightarrow \langle \Phi | \Psi_e \rangle&=&0 ~~ .
\end{eqnarray}
We may now form the basic equations for the excited state, given by
\begin{equation}
\epsilon_e {\cal X}_I^e = \langle \Phi | C_I^- e^{-S} [H,X^e] e^S | \Phi 
\rangle ~~ , \forall I \ne 0 ~~ , 
\label{temp1}
\end{equation}
which is a generalized set of eigenvalue equations with 
eigenvalues $\epsilon_e$
and corresponding eigenvectors ${\cal X}_I^e$. 
We note that the choice of clusters for the excited-state may be different from 
those for the ground state. For example, the  ground state for the Heisenberg 
model on bipartite lattices is in the subspace $s_T^z \equiv \sum_i s_i^z = 0$, whereas 
the excited state has $s_T^z \equiv \sum_i s_i^z = +1$. The number of 
excited-state ``fundamental'' clusters that are distinct under the translational
and point-group symmetries of the lattice and Hamiltonian is given 
by $N_{f_e}$. 

\section{High-Order Excited-State Operators and Commutations}
 
In a similar manner as for the ground-state, we now define
excited state operator via 
\begin{equation}
X^e = \sum_{l} ~ \sum_{i_1,\cdots,i_l} ~ {\cal X}^e_{i_i, \cdot\cdot\cdot, i_{l}}
s^+_{i_1} \cdot\cdot\cdot s^+_{i_{l}}
\label{hoeso1}
\end{equation}
where the indices $\{i_1, \cdots, i_l \}$  run over all lattice sites. 
We assume explicitly again that  there 
are $(l!)$ orderings of the indices (even for $s > \frac 12$). 
The index $I$ corresponds to one of the choices of $\{i_1, \cdots, i_l \}$ for the
fundamental set of configurations for the excited state, such that ${\cal X}^e_I 
\equiv {\cal X}^e_{i_i, \cdot\cdot\cdot, i_{l}}$. 
We now also define the further high-order operators for the excited state, given by
\begin{equation}
~ \left .
\mbox{
\begin{tabular}{l@{~}l@{~}l@{~~}}
$P_{k}$   &$\equiv$ &$\sum_{l} ~ \sum_{i_2,\cdots,i_l]} ~ 
l {\cal X}^e_{k, i_2, \cdot\cdot\cdot, i_{l}} ~
s^+_{i_2} \cdot\cdot\cdot s^+_{i_{l}}$\\
$Q_{km}$  &$\equiv$ &$\sum_{l>1} ~ \sum_{i_3,\cdots,i_l} ~ 
l(l-1){\cal X}^e_{k, m, i_3, \cdot\cdot\cdot, i_{l}} ~
s^+_{i_3} \cdot\cdot\cdot s^+_{i_{l}}$  \\
$R_{kmn}$ &$\equiv$ &$\sum_{l>2} ~ \sum_{i_4,\cdots,i_l} ~ 
l(l-1)(l-2){\cal X}^e_{k, m, n, i_4, \cdot\cdot\cdot, i_{l}} ~
s^+_{i_4} \cdot\cdot\cdot s^+_{i_{l}}$  \\
$T_{kmnp}$&$\equiv$ &$\sum_{l>3} ~ \sum_{i_5,\cdots,i_l} ~
l(l-1)(l-2)(l-3){\cal X}^e_{k,m,n,p, i_5, \cdot\cdot\cdot, i_{l}} ~
s^+_{i_5} \cdot\cdot\cdot s^+_{i_{l}}$ 
\end{tabular}
}
\right \}
\label{hoeso2}
\end{equation}
The following commutation relations may also be proven:
\begin{equation}
~ \left .
\mbox{
\begin{tabular}{l@{~}l@{~}l@{~~}}
$[s_k^{z},X^e]$    ~ &=& ~ $P_{k} s_k^+$~~,\\
$[s_k^{-},X^e]$  ~ &=& ~ $-2P_{k} s_k^z - Q_{kk}s_k^+$~~,\\
$[s_k^{z},P_{m}]$    ~ &=& ~ $Q_{km} s_k^+$~~,\\
$[s_k^{z},Q_{mn}]$  ~ &=& ~ $R_{kmn} s_k^+$~~,\\
$[s_k^{z},P_m^2]$    ~ &=& ~ $2 P_m Q_{km} s_k^+$~~,\\
$[s_k^{-},P_{m}]$    ~ &=& ~ $-2 Q_{km} s_k^z - R_{kkm} s_k^+$~~,\\
$[s_k^{-},P_{m}^2]$  ~ &=& ~ $-2 Q_{km}^2 s_k^+ - 2 P_m R_{kkm} s_k^+
-4P_m Q_{km} s_k^z $~~,\\
$[s_k^{z},R_{mnp}]$  ~ &=& ~ $T_{kmnp} s_k^+$~~,\\
$[s_k^{-},Q_{mn}]$  ~ &=& ~ $-2 R_{kmn} s_k^z - T_{kkmn} s_k^+$~~.\\
\end{tabular}
}
\right \} ~~
\label{hoeso3}
\end{equation}

\section{Deriving and Solving The Excited State Equations}

We now wish to determine and solve the CCM excited-state equations given by Eq. (\ref{temp1}) 
Specific terms in the Hamiltonian are now explicitly written in terms of the new excited-state
high-order CCM operators as:
\begin{eqnarray}
{\rm TERM~1:} ~ e^{-S} [s^z_i s^z_j, X^e] e^S &=& 
P_i s_i^+ s_j^z + P_i F_j s_i^+ s_j^+ + 
P_j s_j^+ s_i^z + P_j F_i s_i^+ s_j^+ + 
Q_{ij} s_i^+ s_j^+ 
\nonumber \\
{\rm TERM~2:} ~e^{-S} [s^z_i s^+_j, X^e] e^S &=& 
P_i s_i^+ s_j^+ 
\nonumber \\
{\rm TERM~3:} ~e^{-S} [s^z_i s^-_j, X^e] e^S  &=& 
-2 P_i F_j s_i^+ s_j^z - P_i G_{jj} s_i^+ s_j^+ - P_i F_{j}^2  s_i^+ s_j^+ -2 P_{j}  s_i^z s_j^z 
-2 P_j F_j s^+_j  s^z_i \nonumber \\
&&-2P_j F_i s_i^+ s_j^z  - 2 P_j G_{ij} s_i^+ s_j^+ - 2 P_j F_i F_j s_i^+ s_j^+ 
-2Q_{ij} s_i^+ s_j^z
\nonumber \\
&&-2Q_{ij} F_j s_i^+ s_j^+-Q_{jj} s_j^+ s_i^z - Q_{jj} F_i s_i^+ s_j^+ - R_{ijj} s_i^+ s_j^+
\nonumber \\
{\rm TERM~4:} ~ e^{-S} [s^+_i s^z_j, X^e] e^S  &=& 
P_j s_i^+ s_j^+ 
\nonumber \\
{\rm TERM~5:} ~e^{-S} [s^-_i s^z_j, X^e] e^S  &=& 
-2 P_j F_i s_j^+ s_i^z - P_j G_{ii} s_i^+ s_j^+ - P_j F_{i}^2  s_j^+ s_i^+ -2 P_{i}  s_i^z s_j^z 
-2 P_i F_j s^+_j  s^z_i \nonumber \\
&&-2P_i F_i s_i^+ s_j^z  - 2 P_i G_{ij} s_i^+ s_j^+ - 2 P_i F_i F_j s_i^+ s_j^+ 
-2 Q_{ij} s_j^+ s_i^z
\nonumber \\
&&-2Q_{ij} F_i s_i^+ s_j^+ - Q_{ii} s_i^+ s_j^z - Q_{ii} F_j s_i^+ s_j^+ - R_{iij} s_i^+ s_j^+
\nonumber \\
{\rm TERM~6:} ~e^{-S} [s^+_i s^-_j, X^e] e^S &=& 
-2 P_j s_i^+ s_j^z - Q_{jj} s_i^+ s_j^+ -2P_j F_j s_i^+ s_j^+
\nonumber \\
{\rm TERM~7:} ~e^{-S} [s^-_i s^+_j, X^e] e^S &=& 
-2 P_i s_j^+ s_i^z - Q_{ii} s_i^+ s_j^+ -2P_i F_i s_i^+ s_j^+
\nonumber \\
{\rm TERM~8:} ~e^{-S} [s^+_i s^+_j, X^e] e^S  &=& 0
\nonumber \\
{\rm TERM~9:} ~ e^{-S} [s^-_i s^-_j, X^e] e^S &=&
4 P_i F_j s_i^z s_j^z +4P_iG_{ij} s_i^+ s_j^z +2P_i G_{jj} s_j^+ s_i^z +2 P_i M_{ijj} s_i^+ s_j^+ 
\nonumber \\ && +4P_iF_iF_js_i^+s_j^z +4P_iF_j G_{ij} s_i^+ s_j^+ + 2 P_i F_i G_{jj} s_i^+  s_j^+ 
+2P_iF_i F_j^2 s_i^+ s_j^+ 
\nonumber \\ &&+2P_i F_j^2 s_j^+ s_i^z + 2Q_{ii} F_j s_i^+ s_j^z +Q_{ii} G_{jj} s_i^+ s_j^+ 
+ Q_{ii} F_j^2 s_i^+ s_j^+ 
\nonumber \\ &&+4 Q_{ij} s_i^z s_j^z + 4Q_{ij} F_j s_j^+ s_i^z + 4 Q_{ij} F_i s_i^+ s_j^z  
+4Q_{ij} G_{ij} s_i^+ s_j^+ 
\nonumber \\ &&+4 Q_{ij}F_i F_js_i^+ s_j^+ + 2R_{iij} s_i^+ s_j^z + 2R_{iij} F_j s_i^+ s_j^+
+ 2R_{ijj} s_j^+ s_i^z 
\nonumber \\ && +2 R_{ijj} F_i s_i^+ s_j^+ + T_{iijj} s_i^+ s_j^+ + 4 P_j G_{ij} s_j^+ s_i^z 
+ 2  P_j M_{iij} s_i^+ s_j^+ 
\nonumber \\ &&+ 4 P_j F_i s_i^z s_j^z +4 P_j F_i F_j s_j^+  s_i^z + 2P_j G_{ii} s_i^+ s_j^z 
+ 2P_j G_{ii} F_j s_i^+ s_j^+ 
\nonumber \\ && +2P_j F_i^2 s_i^+ s_j^z + 2 P_j F_i^2 F_j s_i^+ s_j^+ +4P_j F_i G_{ij} s_i^+ s_j^+
+ 2 Q_{jj} F_i s_j^+ s_i^z 
\nonumber \\ && +Q_{jj}  G_{ii} s_i^+ s_j^+ + Q_{jj} F_i^2 s_i^+ s_j^+ 
\nonumber \\
{\rm TERM~10:} ~ e^{-S} [s^z_i , X^e] e^S &=& 
P_i s_i^+
\nonumber \\
{\rm TERM~11:} ~e^{-S} [s^-_i, X^e] e^S &=& 
-2 P_i s_i^z - Q_{ii} s_i^+- 2P_i F_i s_i^+
\nonumber \\
{\rm TERM~12:} ~ e^{-S} [(s^z_i)^2, X^e] e^S &=& 
2P_i s_i^+ s_i^z + Q_{ii} (s_i^+)^2 + P_i (s_i^+)^2 + 2 P_i F_i (s_i^+)^2 
\nonumber \\
{\rm TERM~13:} ~e^{-S} [s^+_i, X^e] e^S &=& 0
\label{hoese4}
\end{eqnarray}
(Note that $s^-|\Phi\rangle=0$ is again implicitly assumed in Eq. (\ref{hoese4}) above.) 
Again, we now ``pattern-match''  the $C_I^-$ operators (this time with respect to the fundamental 
set of the clusters in the excited state) to those the relevant terms in the 
Hamiltonian from Eq. (\ref{temp1}) above in order to form the CCM excited-state equations 
at a given level of approximation. By contrast to the case for the ground state, we see that the 
high-order operators of Eq. (\ref{hoeso2}) are in linear in those terms in Eq. (\ref{hoese4}). 
We choose the eigenvalue of lowest value to be our result, and this method was found
to provide good results in regions of the parameter space for which the model state was
a good choice. 
Again we note that we have formed an eigenvalue problem, which is readily
solved using a standard eigenvalue solver. However, the computational problem thus formed 
uses local memory that scales
with the number of fundamental clusters used in the excited state, i.e., as $N_{f_e}^2$. 
%Again, this becomes prohibitive computationally for extremely large values of $N_{f_e}$ and
%so we again use direct iteration methods.

%\section{Direct Iteration of the Excited-State Equations and Parallelization}

The eigenvalue equations of Eq. (\ref{temp1}) may be iterated directly in order to solve
them. We denote the matrix for the eigenvalue problem of Eq. (\ref{temp1}) by $B$ and we
denote the eigenvectors by $y=({\cal X}_1^e, \cdots , {\cal X}^e_{N_{f_e}})^T$. Hence, we iterate
directly the eigenvalue equation given by
\begin{equation}
B y = \lambda y ~~ .
\label{hoese5} 
\end{equation}
This is just the well-known ``power iteration'' method and the ratios of
$X^e_I$ in successive iterations yields the relevant eigenvalue.  However,
the eigenvalue determined in this manner is the eigenvalue of largest magnitude, 
$\lambda_{\rm MAX}$, rather than the lowest (generally the one of smallest magnitude 
$\lambda_{\rm MIN}$  for our purposes) 
that we wish to obtain here. Thus, we use find the eigenvalue of smallest magnitude 
by using the ``shifted'' power iteration method. Once $\lambda_{\rm MAX}$ has been found,
we then solve the following eigenvalue equation by direct iteration:
\begin{equation}
(B - \lambda_{\rm MAX}I) y' = \lambda' y' ~~. 
\label{hoese6}
\end{equation}
This process ought to converge to an eigenvalue $\lambda'=\lambda_{\rm MIN}-
\lambda_{\rm MAX}$. Indeed, this was found to be the case for those spin models 
for which the model state was a ``good choice''. Indeed, the lowest-valued
eigenvalue obtained in this manner agreed perfectly with those results for the
eigenvalue of lowest values obtained via a complete diagonalization of the matrix 
eigenvalue problem at every level of approximation. 

\section{Extrapolation of Expectation Values}

In practice, we often need to extrapolate individual LSUB$m$ or SUB$m$-$m$ (etc.)  
expectation values  $y(m)$ in the limit $m \rightarrow \infty$. Indeed, the extrapolation of CCM expectation values is pivotal to its practical use for quantum magnetic systems.  However, there 
are many aspects of such extrapolations  that we still do not understand and that have not 
been tested. Furthermore, the extrapolation of CCM expectation values poses two distinct
problems: a) no provable exact rules of extrapolation are known (unlike, e.g., 
finite-size exact diagonalisations); and, b) we have only small numbers of data points (i.e.,
7 or 8 at most for 1D systems).  Hence, until now, only heuristic or ``ad hoc'' schemes have been 
used. These heuristic schemes are ``parametric'' in the sense that the data is fitted to 
specific functions (e.g., polynomials) that are defined prior to fitting the data and where
the coefficients of these functions are the basic ``parameters.'' However, one should note 
that the results of such parametric/ad hoc extrapolation 
procedures have been shown to yield consistently valid results for an extremely wide range of 
quantum spin systems, see, e.g., Refs. \cite{ccm12,ccm15,ccm20}. Indeed, the use of
such schemes has been shown to provide improved results in all cases where the 
model state might even be suspected to be a reasonable starting point.  This is a very strong 
vindication of the use of these simple extrapolation procedures. 

We now turn to specific 
instances of such extrapolation schemes, and we start by noting that a common 
scheme for extrapolating the ground-state energy is given by
\begin{equation}
y(m)_{\rm SCHEME~1}=a_0+a_1m^{-2}+b_2m^{-4} ~~. 
\label{scheme1}
\end{equation}
Thus far, least-squares fits of the data to this scaling rule have been performed, 
where the extrapolated value 
$y(m \rightarrow \infty)_{\rm SCHEME~1}$ is given by $a_1$. For other expectation values, 
such as the sublattice magnetisation and excitation energy gap,  a common scheme is given by
\begin{equation}
y(m)_{\rm SCHEME~2}=b_0+b_1 m^{-1}+b_2m^{-2} ~~.
\label{scheme2}
\end{equation}
In this case,  the extrapolated value $y(m \rightarrow \infty)_{\rm SCHEME~2}$ is given by 
$b_0$. Finally, another common scheme is given by
\begin{equation}
y(m)_{\rm SCHEME~3}=c_0+c_1 m^{c_2} ~~.
\label{scheme3}
\end{equation}
The extrapolated value $y(m \rightarrow \infty)_{\rm SCHEME~3}$ is given by 
$c_0$. Other schemes have included Pad\'e approximants \cite{ccm15} and similar schemes
to Eqs. (\ref{scheme1}) and (\ref{scheme2}) for fixed, though non-integer, exponents
\cite{ccm31,ccm32,ccm33,ccm34} have been used. For example, the sublattice magnetisation in Refs. \cite{ccm31,ccm32,ccm33,ccm34} was found to scale as: $y(m)=d_0+d_1m^{-0.5}+d_2m^{-1.5}$. 

In future, we would ideally like to establish exact rules of scaling of expectation 
values with approximation level and/or go to much higher orders of approximation 
so that we have more data points to extrapolate.
Evidence for exact rules of scaling is supported by the fact that the ground-state
energy of the Heisenberg model on bipartite lattices such as the linear chain 
and square lattice appear to follow the scheme of Eq. (\ref{scheme1}) quite well. 
By contrast, their magnetisations and excitation energy gaps appear to follow the 
scheme of Eq. (\ref{scheme2}). Few other general rules seem to exist and
certainly no such rules have, as yet, been proven mathematically. Furthermore, we
are must use computationally intensive methods in parallel to solve for levels 
of approximation currently available.  Further increases in approximation level 
might be possible in future due to computational improvements, although we might 
still be restricted to fairly small numbers of data points for the current approximation 
schemes. Hence, both of these goals may be not be achieveable in practice. 
In the absence of either of these goals, one might wish to extrapolate using a variety of 
extrapolation schemes in order to determine (in broad terms only) the amount of
their mutual agreement. Furthermore, odd and even series of (e.g., LSUB$m$ 
expectation values) ought to extrapolate to the same value (see Ref. \cite{ccm36}). 
Again, this might yield a rough idea of errors of extrapolation. 

Future research might also concentrate on the  
establishment of new approximation schemes that do not scale as quickly
as LSUB$m$ and SUB$m$-$m$ schemes and yet still appear scalable with 
approximation level. However, it is likely that the number of fundamental clusters 
in any such scheme will still increase exponentially with approximation level.
It would be beneficial to carry out a  fully ``statistical'' exploration of the topic of 
parametric extrapolations such as those of Eqs. (\ref{scheme1}) to (\ref{scheme3}) 
based on only small numbers of data points. This might yield better
extrapolations, but perhaps more importantly it would hope to establish 
with more mathematical rigour the estimate of the error in the extrapolated values.  
Until now, the extrapolated figures have generally been presented with no estimated 
degree of error. Furthermore, this ``statistical'' approach might also include an exploration 
of, e.g., least-squares,  weighted least-square and maximum likelihood methods fitting 
methods of the parametric scaling laws  to the data and the effect that this would have 
on the extrapolated results. Indeed, hitherto, only ad hoc/parametric models or rules  
of the types shown in Eqs. (\ref{scheme1}) to (\ref{scheme3}) have been 
used to carry out extrapolation of CCM data using least-squared methods. 
Finally, a outline of  ``best practice'' for extrapolating CCM data (i.e., which extrapolation 
schemes and methods to use and exactly what to report) for given types of 
expectation values might prove very useful.  These topics remain for future study however.

%We note that the local memory usage is again far lower for this ``power iteration'' 
%approach rather than the corresponding ``complete diagonalization'' of the matrix 
%problem because we need store only the values for the set $\{{\cal X}^e_I\}$ at each point. 
%Finally, the direct iterative solution of the CCM excited-state problem may be parallelized 
%readily. Again, we share the problem of finding the left-hand side of Eq. (\ref{hoese5}) 
%over all processors for different values of $I$. We then collect the results 
%together in order to form the right-hand side of Eq. (\ref{hoese5}). We iterate 
%to find $\lambda_{\rm MAX}$. A similar parallelization process is then used to find 
%$\lambda_{\rm MIN}$.



\end{document}